\documentclass[prl,twocolumn,showpacs]{revtex4}
\usepackage{amsfonts,amsmath,mathrsfs,epsfig,amsbsy,bm,verbatim,subfigure}

\newcommand{\ua}{\uparrow}
\newcommand{\da}{\downarrow}

\def\ga{{\ \lower-1.2pt\vbox{\hbox{\rlap{$>$}\lower5pt\vbox{\hbox{$\sim$}}}}\ }}
\def\la{{\ \lower-1.2pt\vbox{\hbox{\rlap{$<$}\lower5pt\vbox{\hbox{$\sim$}}}}\ }}

\def\ua{\uparrow}
\def\da{\downarrow}

\def\beq{\begin{equation}}
\def\eeq{\end{equation}}

\def\bea{\begin{eqnarray}}
\def\eea{\end{eqnarray}}

\begin{document}
\title{A Majorana smoking gun for the superconductor-semiconductor hybrid topological system}
\author{S. Das Sarma$^1$}
\author{Jay D. Sau$^2$}
\author{Tudor D. Stanescu$^3$}
\affiliation{
$^1$Condensed Matter Theory Center, Department of Physics, University of Maryland, College Park, Maryland 20742-4111, USA\\
$^2$Department of Physics, Harvard University, Cambridge, Massachusetts 02138, USA\\
$^3$Department of Physics, West Virginia University, Morgantown, WV 26506,USA}

\begin{abstract}
Recent observations of a zero bias conductance peak in tunneling transport measurements in superconductor--semiconductor nanowire devices provide evidence for the predicted  zero--energy Majorana modes, but not the conclusive proof for their existence. We establish that direct observation of a splitting of the zero bias conductance peak can serve as the smoking gun evidence for the existence of the Majorana mode. We show that the splitting has an oscillatory dependence on the Zeeman field  (chemical potential) at fixed chemical potential (Zeeman field). By contrast, when the density is constant rather than the chemical potential -- the likely situation in the current experimental set-ups -- the splitting oscillations are generically suppressed. Our theory  predicts the conditions under which the splitting oscillations can serve as the smoking gun for the experimental confirmation of the elusive Majorana mode.
\end{abstract}

\maketitle
The recent experimental report \cite{Mourik} providing direct observational evidence for the possible existence of the predicted \cite{Sau,Long-PRB,Roman,Oreg} zero--energy Majorana quasiparticle in superconductor--semiconductor nanowire  hybrid structures in the presence of 
 spin-orbit coupling and Zeeman splitting has tremendously excited the whole physics community \cite{Reich,Wilczek,Brouwer_science,Physicstoday}. Yet, in spite of several subsequent  reports \cite{Deng, Weizman,Marcus}  having validated the original data of Ref. ~\onlinecite{Mourik}, this experiment has also raised many questions. Most of these questions arise from a critical comparison between the experimental data  \cite{Mourik} and the
original theoretical predictions \cite{Sau,Long-PRB,Roman,Oreg}, leading  to the inevitable conclusion that there are some significant discrepancies between experiment and theory.  For example, the key experimental observation is the development of a robust subgap zero bias conductance peak (ZBCP) in the tunneling differential conductance of the nanowire in the presence of an external magnetic field applied along the wire, as predicted theoretically \cite{Long-PRB} and as expected
for a Majorana zero energy mode in a topological superconductor \cite{Sengupta-2001,R1}. However,  the actual magnitude of the ZBCP ($\sim 0.1e^2/h$) is more than an order of magnitude smaller than the predicted ideal quantized value
$(2e^2/h)$.  In addition, the Majorana--induced ZBCP should only appear beyond a magnetic field--driven topological quantum phase transition (TQPT) characterized by the closing of the superconducting (SC) gap, yet  there is no apparent signature of gap closing in the measured tunneling current. Although recent theoretical works \cite{Lin2012,Prada,Stanescu2012}
 provide reasonable explanations for some of these discrepancies, other recent papers emphasize that a ZBCP could arise in the system in the absence of Majorana bound states, due to more mundane mechanisms involving strong disorder \cite{Liu, Altland, Beenakker_Weak}, smooth end confinement \cite{Kells}, or Kondo physics \cite{Silvano}.

Given this fluid and  confusing nature of the subject matter, with publications arguing in favor of or against the Majorana interpretation of the experimental observation in Ref. \onlinecite{Mourik} appearing almost weekly,   it is of paramount importance to conceive of hallmark experimental signatures for the Majorana quasiparticle.  It was already emphasized in the original theoretical predictions \cite{Roman,Long-PRB} that the observation of a ZBCP at finite magnetic field is only a necessary condition for the existence of Majorana quasiparticles. The sufficient condition to validate their existence must be some type of interference measurement, such as the fractional Josephson effect \cite{Kwon,kitaev} manifesting a $4\pi$ periodicity in an ac Josephson measurement. While such a measurement, or the direct observation of non-Abelian Majorana interference \cite{alicea,sau_int,2D_int}, will certainly be necessary down the line to absolutely validate the existence of localized non-Abelian Majorana modes, the high level of complexity and difficulty of this type of measurements make it unlikely that they will be successful in the near future.   Therefore, it is desirable that a simpler experimental smoking gun for the Majorana, something with a difficulty level comparable with to the existing ZBCP experiments, be proposed and carried out long before the rather challenging fractional Josephson measurement and the interference experiments could be performed. 
In the current work, we propose a smoking gun Majorana measurement that could not only be carried out right now, but in fact it is conceivable that the necessary experimental data could already be hidden in the reported ZBCP measurements.

Our key observations in this context are: i) nanowire Majorana modes always come in pairs \cite{kitaev} localized at the two ends of the wire, and ii) the Majorana mode is a pure zero--energy  mode only when the wire is infinitely long. For any realistic finite--length wire, the two end Majorana wavefunctions overlap and the resulting hybridization leads to a splitting of the zero-mode \cite{meng}. This hybridization--induced energy--splitting increases with either decreasing wire length or increasing  coherence length and is characterized by an oscillatory behavior determined  by the Fermi wavevector of the top occupied band \cite{meng}. This is in contrast to other sources of zero-bias conductance
peaks such  as Kondo resonances \cite{Silvano} where the Zeeman splitting is expected to produce a monotonically increasing splitting. Since both the effective SC coherence length and the effective Fermi wavevector can be tuned by varying the external magnetic field or the chemical potential of the system, a direct observation of the oscillatory energy splitting  with increasing magnetic field (which increases the coherence length by increasing the effective Fermi-velocity and suppressing the SC gap) should be a definitive smoking gun for the Majorana existence.  

In real systems, there are two main challenges that have to be addressed: the finite energy resolution and the limited experimental ability to control the chemical potential. Various broadening mechanisms (e.g., temperature,
inelastic scattering, quasiparticle poisoning, disorder, finite tunnel barrier) may mask the underlying Majorana splitting oscillations as two closely--spaced split peaks near zero bias may merge into
a single broad zero--bias peak.  Nonetheless, the width of the Majorana peak should show a  modulation with varying Zeeman field or chemical potential, 
even if the finite resolution masks the splitting itself.
  In the current work we make detailed theoretical predictions about how the Majorana ZBCP splitting
 should  depend on the relevant parameters. These predictions are experimentally verifiable and can serve as a clear smoking gun for the validation (or invalidation) of the observation of Majorana modes in hybrid nanostructures. Since four distinct experimental groups \cite{Mourik,Weizman,Deng,Marcus} have already reported detailed results for the ZBCP dependence on  the applied Zeeman field, it appears that our proposed Majorana smoking gun measurement should be feasible in the very near future.

The currently existing experimental results do not manifest any clear signature of Majorana oscillatory splitting, although the issue is by no means definitively sorted out since the higher field data are rather sparse and have not yet been carefully analyzed. 
However, more importantly, our work shows that one must conceptually distinguish between the field--dependent Majorana
splitting at constant chemical potential versus constant carrier density.   All discussions of  Majorana splitting in the literature have so far considered only the constant chemical potential situation,  where both the average splitting and the oscillation amplitude increase monotonically with increasing Zeeman splitting, yet actual experiments, because of repulsive interactions are mostly carried out at constant density.
We show that the constant density condition is qualitatively different and may lead to a strong suppression of the oscillatory splitting, although generically some splitting  is always present.  Moreover, the oscillatory splitting is always present at constant Zeeman field as a function of chemical potential (or particle density).  Thus, experiments searching for the oscillatory splitting  (or the associated modulations of the ZBCP width) in gated samples should focus on the ability to tune the density and/or chemical potential.  We know of no other proposed mechanism in superconductor--semiconductor hybrid structures that could lead to a ZBCP manifesting oscillatory splitting, except for the Majorana--induced zero bias peak predicted to occur in the topological SC phase. 

\begin{figure}[tbp]
\begin{center}
\includegraphics[width=0.48\textwidth]{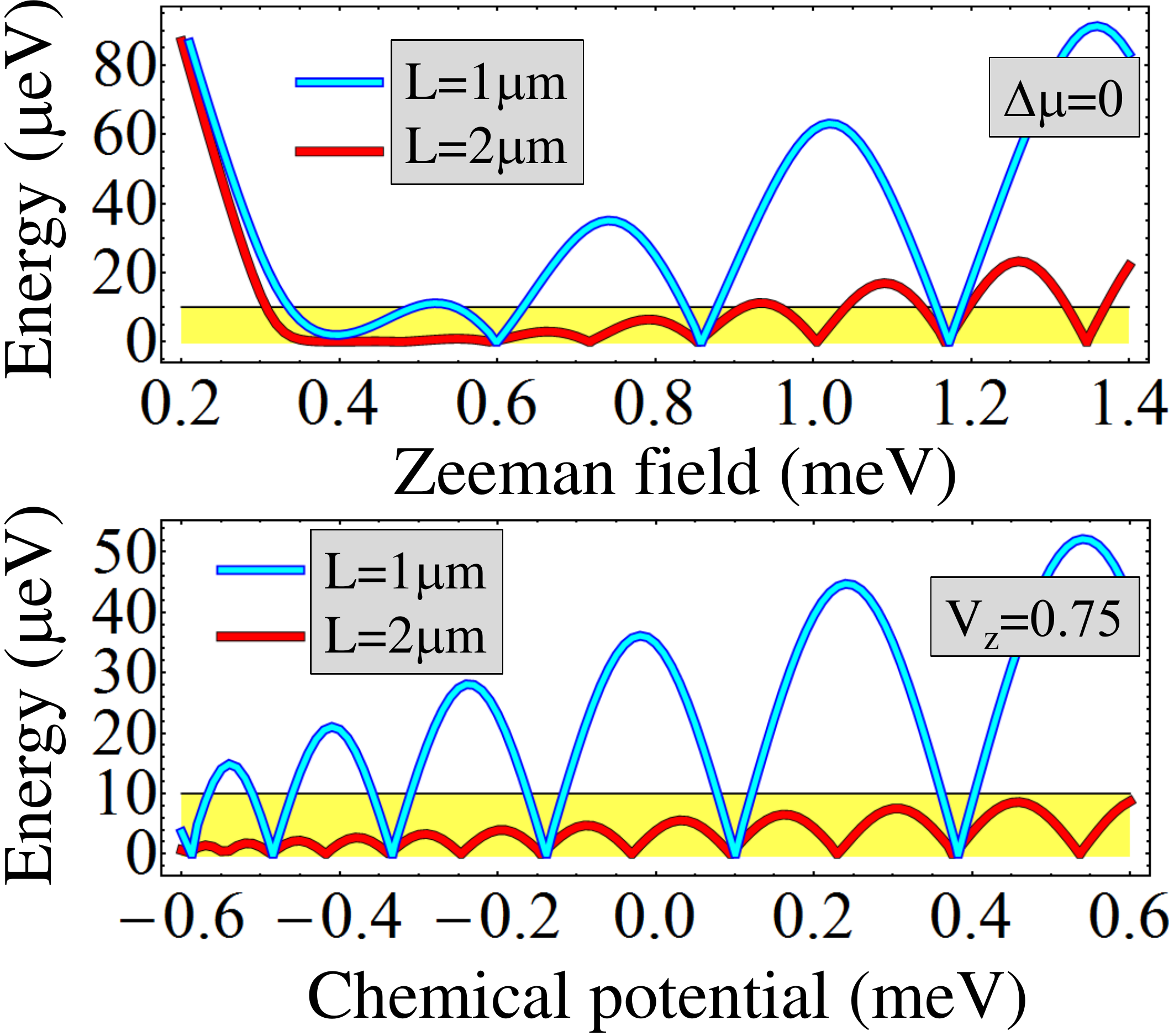}
\vspace{-7mm}
\end{center}
\caption{(Color online) Oscillatory splitting of the Majorana mode as a function of $V_Z$ (for $\mu=\mu_3$, top panel) and 
$\Delta\mu=\mu-\mu_3$ (for $V_Z=0.75$ meV, bottom). The parameters used in the calculation are:  $\Delta=0.25$ meV, $\alpha^2 m^*/2\hbar^2=50\mu eV$, and $m_{eff}=0.015 m_e$. This behavior is independent on the number of occupied bands. Finite energy resolution ($\delta E=10\mu$eV, yellow band) may prevent the observation of oscillatory slitting in long wires. }
\vspace{-6mm}
\label{Fig1}
\end{figure}

Finite SC nanowires in the topological phase have been proposed to support a pair of near zero-energy Majorana fermions (MFs), one at each end of the wire. The energy degeneracy of these zero modes is removed \cite{meng} by an energy splitting
\begin{equation}
\Delta E \sim Re[\Psi_l^\dagger(L/2)\hat{J}\Psi_r(L/2)],
\end{equation}  
where $\Psi_{l,r}(L/2)$ are the wave-functions of the MF bound-states near the left and right ends respectively and $\hat{J}$ 
is the splitting operator whose explicit form is given in the Supplementary Material.
The spatial dependence of the MF wave-functions, which can be approximated as (see Supplementary Material for details)
\begin{align}
&\Psi_{l}(x)\propto e^{-x/\xi}e^{\pm i k_{F,eff}x},\label{eq:Psi}
\end{align}
for $x\gg \xi$ where $\xi$ is the effective coherence length and $k_{F,eff}$ is the effective 
fermi-wave-vector associated with the zero-mode solution. The parameter $k_{F,eff}$ and $\xi$ 
depend on the microscopic parameters such as the Zeeman splitting $V_Z$ and the chemical $\mu$, 
which can be tuned externally by changing the applied in-plane magnetic field or 
gate voltage. The wave-function $\Psi_r(x)$ of the MF at the right end behaves in a qualitatively 
similar way, so that for $L\gg \xi$ the MF splitting wave-function has the approximate form 
\begin{align}
&\Delta E\approx \hbar^2 k_{F,eff}\frac{e^{-2 L/\xi}}{m\xi}\cos{(k_{F,eff} L)},\label{eq:simple}
\end{align}
where $m$ is the effective electron mass in the nanowire. The $\cos{(k_{F,eff}L)}$ factor should lead to an oscillation  of the energy splitting $\Delta E$ as a function of the separation $L$ between the MF. In the nanowire  case, it is more natural to tune the Zeeman potential $V_Z$ or the gate voltage $\mu$. Since $k_{F,eff}$ 
depends on $V_Z$ and $\mu$, for sufficiently long $L$, one can expect the $\cos{\{ k_{F,eff}(V_Z,\mu)L\}}$ 
factor to lead to oscillations in $\Delta E$ as a function of both $V_Z$ and $\mu$.
We note that the $e^{-2L/\xi}$ factor leads to the well-known exponential topological protection of the zero-energy MF,
 which, however, only applies in the very long-wire limit of $L\gg\xi$ where Eq. (3) is valid.

\begin{figure}[tbp]
\begin{center}
\includegraphics[width=0.48\textwidth]{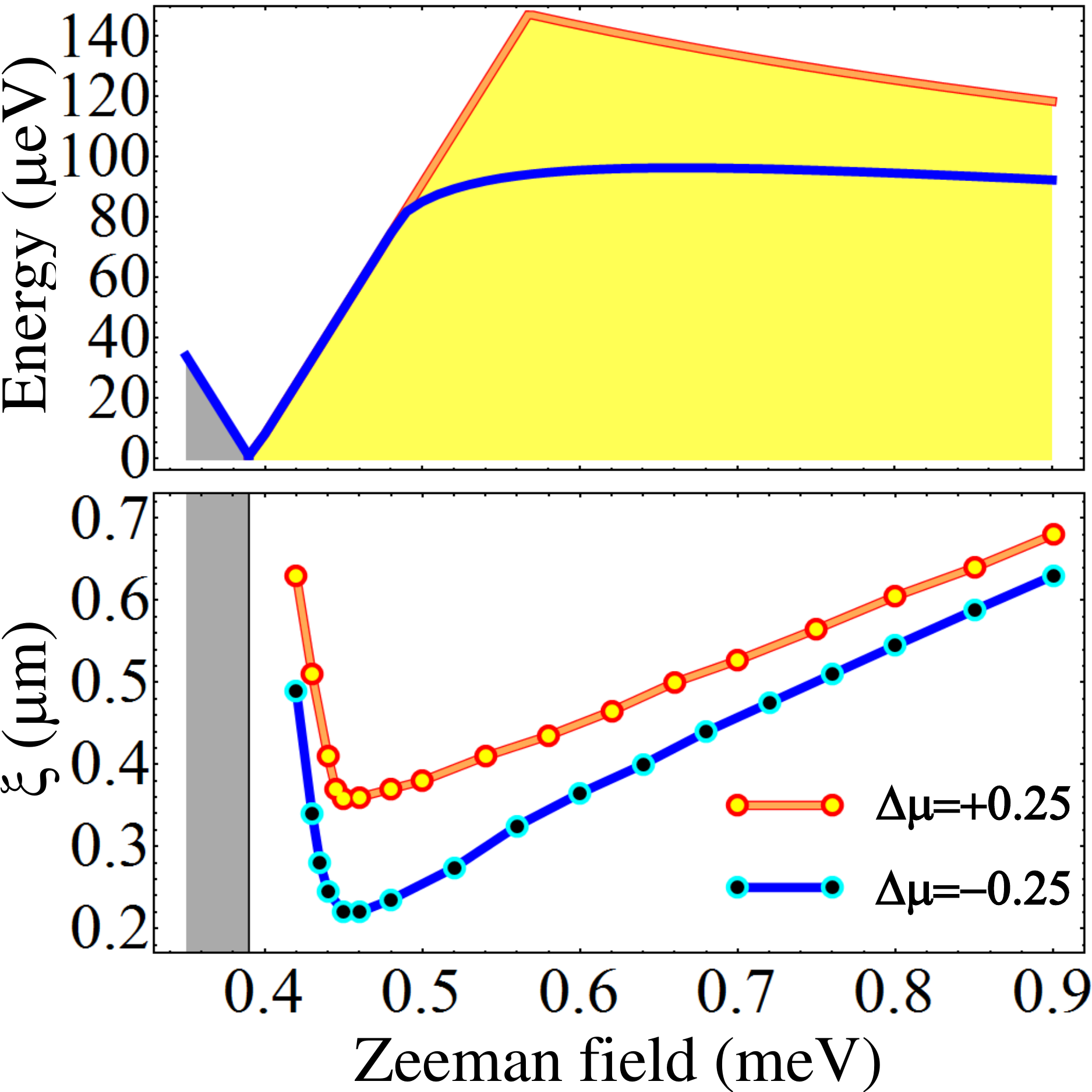}
\vspace{-7mm}
\end{center}
\caption{(Color online) Top: Dependence of the SC gap on the Zeeman field in the topological SC phase (yellow) for two different values of the chemical potential. Bottom: Decay length of the MFs as a function of Zeeman splitting. Note that the minimum of the coherence length occurs at $V_Z\gtrsim V_{Z,c}$, right above the critical field, and does not coincide with the maximum of the energy gap.}
\vspace{-6mm}
\label{Fig2}
\end{figure}

To confirm these expectations,  we study the spectrum of a spin--orbit coupled 
semiconductor nanowire in proximity to a superconductor and placed in a magnetic field parallel to the wire using the 
numerical diagonalization of the Bogoliubov-de Gennes (BdG) Hamiltonian \cite{SLDS}.
 Since we are considering  an experimentally relevant quasi--1D nanowire (with diameter $\sim 80$nm), our system can support 
 several sub--bands with dispersion minima near the spin--degenerate energies $E_n=E_n(k=0, V_z=0)$. As the energy separation between the sub--bands is larger than the SC gap, the top--most  occupied band $n_{\rm top}$, with effective chemical potential $\Delta\mu=\mu-\mu_{n_{\rm top}}$, where $\mu_{n_{\rm top}}=E_{n_{\rm top}}$,  dominates the low energy part of the spectrum, including the properties of the MF bound states.  
The splitting of the MFs for a nanowire with $n_{\rm top}=3$  as function of $V_Z$ and $\Delta \mu=\mu-\mu_3$ is shown in Fig.~\ref{Fig1} for two different realistic 
wire lengths. As expected from Eq.~\ref{eq:simple}, the energy splitting $\Delta E$ is oscillatory and the oscillation period becomes shorter in longer nanowires, while the amplitude of the oscillations decreases. 
The exponential decrease of the oscillation amplitude is controlled by the factor $e^{-L/\xi}$ in Eq.~\ref{eq:simple}, which reflects the exponential decay of the MF wave--function away from the wire-ends as in Eq.~\ref{eq:Psi}.  This suppression of the energy splitting in long wires depends critically on the actual magnitude of the coherence length $\xi$. 
As seen in Fig.~\ref{Fig2}(b), the calculated coherence length $\xi$ reaches a minimum for a value of $V_Z$ near the 
critical Zeeman field $V_Z\gtrsim V_{Z,c}$,  then increases monotonically as a function of Zeeman field. The increase in 
the coherence length $\xi$ can be attributed to a general decrease in the gap (shown in Fig.~\ref{Fig2}(a)) and a simultaneous increase of the 
fermi velocity with increasing field. In addition, we find a weak increase of the coherence length with $\Delta\mu$. This dependence of the coherence length on $V_Z$ and $\Delta\mu$ is ultimately responsible for the increase of the oscillation amplitude shown in Fig.~\ref{Fig1}. 

What are the main challenges facing the experimental detection of this unique signature of the Majorana mode, the oscillatory splitting of the ZBCP?
First, the broadening of the ZBCP implies finite energy resolution, hence in long enough wires, where the splitting is small, the oscillatory splitting can be resolved only above a certain minimum value of the Zeeman field. For example, in the case of a ZBCP with a broadening of $10\,\mu$eV (shown as the yellow band in Fig.~\ref{Fig1}), the oscillations of the Majorana mode become visible in a  $\sim 1\,\mu$m wire for $V_Z>0.5$ meV. The required Zeeman field for a longer wire ($L\sim 2\,\mu$m) is significantly  larger (see Fig.~\ref{Fig1}). However, reducing too much the length of the wire (to enhance the amplitude of the splitting) results in oscillations with very long period, which might again prevent the observation of the oscillatory behavior. Second, the oscillatory splitting of the Majorana mode as function of $V_z$ is a necessary feature in a system with fixed chemical potential. However, this condition may not be satisfied in the current experiments. We emphasize that, while there appears to be a splitting of the ZBCP in the recent  measurements \cite{Mourik}, the are no obvious oscillations of this splitting. Here we provide a mechanism 
which may lead to a considerable suppression of the splitting oscillations as a function of the magnetic field.

\begin{figure}[tbp]
\begin{center}
\includegraphics[width=0.48\textwidth]{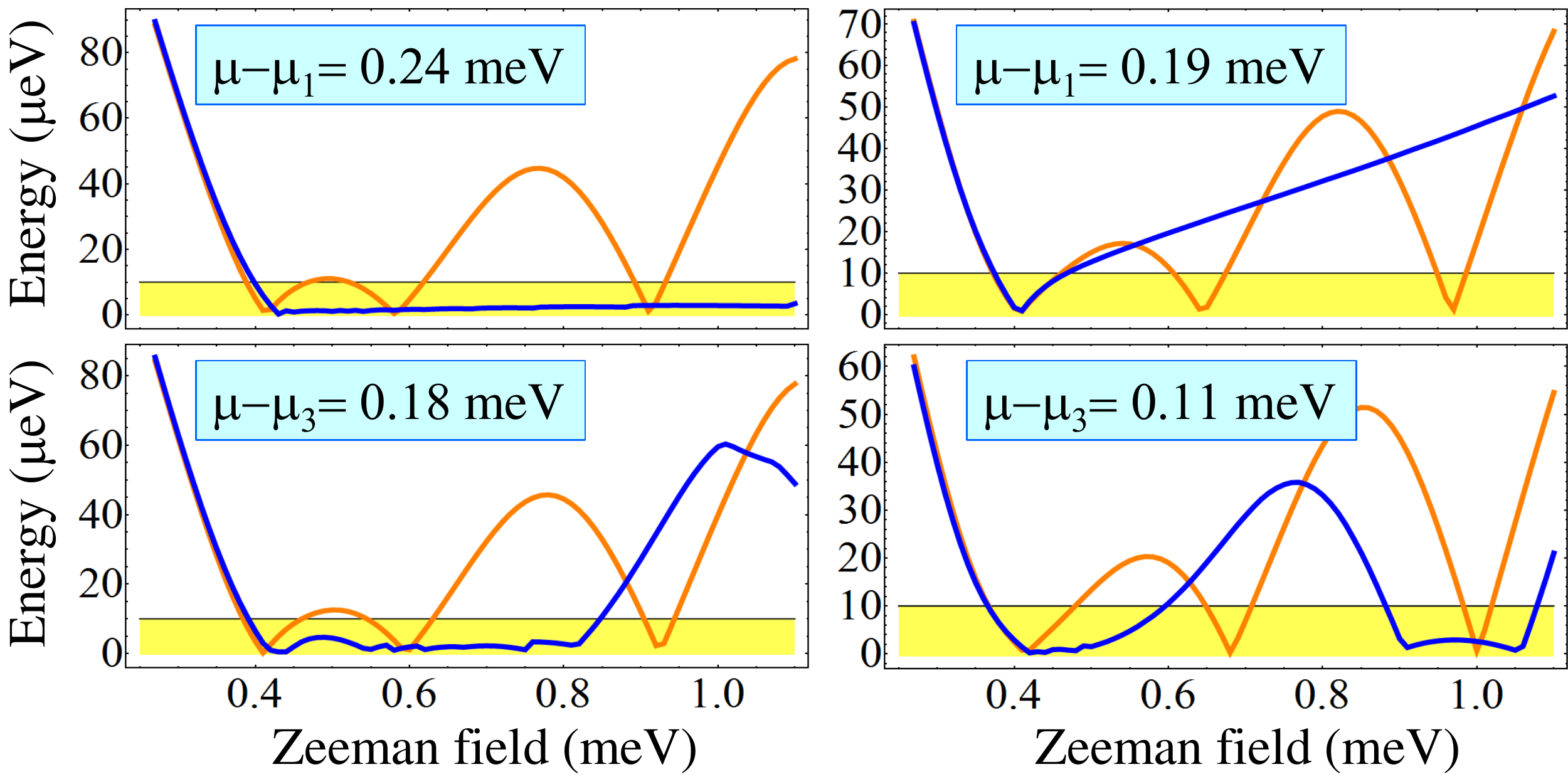}
\vspace{-7mm}
\end{center}
\caption{(Color online) Comparison between the energy splitting of the Majorana mode at constant chemical potential (orange lines) and constant density (blue lines). The quasi--periodicity of the oscillations at constant $\mu$ is absent in a wire with fixed number of particles. The difference is particularly striking in the single--band case (top panels). Under the constant density condition, small variations of the initial occupancy result in qualitative changes of the dependence of $\Delta E$ on $V_Z$.  In contrast, the constant chemical potential condition is characterized by a generic oscillatory splitting. 
}
\vspace{-6mm}
\label{Fig3}
\end{figure}

The basis of our mechanism  is that the Coulomb repulsion among the carriers in the 
nanowire, even after screening by the superconductor, renormalizes the electrostatic potential 
profile in the nanowire as a result of a self-consistent change in the density of electrons. Because of the  repulsive sign of the Coulomb interaction, this extra electrostatic potential will oppose any change of the electron density that might result from variations of the Zeeman potential $V_Z$. Considering, for simplicity,  the extreme limit of strong Coulomb interaction, we obtain the condition that the total number of electrons in the wire mjust be fixed, i.e. the nanowire behaves as a 'constant density' rather than a 'constant chemical potential' system. In the numerical calculations we impose this condition by self--consistently adjusting the chemical potential so that the total number of electrons be the same as the particle number at the TQPT for a nominal non--interacting chemical potential $\Delta \mu_0$. The results are shown in Fig.~\ref{Fig3}. The striking difference between the dependence of the splitting on $V_Z$ at constant chemical potential versus constant density is particularly evident in the single--band case (top panels).  To understand the suppression of the oscillations in the constant density case, we note that a single--band system in the topological phase has one occupied spin-channel. Consequently, the  electron--density in this regime is tied to the Fermi wave---vector $k_{F,eff}\approx n$. 
Since $k_{F,eff}$ dominates the rapidly oscillating part of $\Delta E$ in Eq.~\ref{eq:simple}, if $\mu$ self-consistently adjusts to $V_Z$ 
in a way to keep $k_{F,eff}$ constant, then  $\Delta E$ becomes slowly varying in $V_Z$ as well.
\begin{figure}[tbp]
\begin{center}
\includegraphics[width=0.48\textwidth]{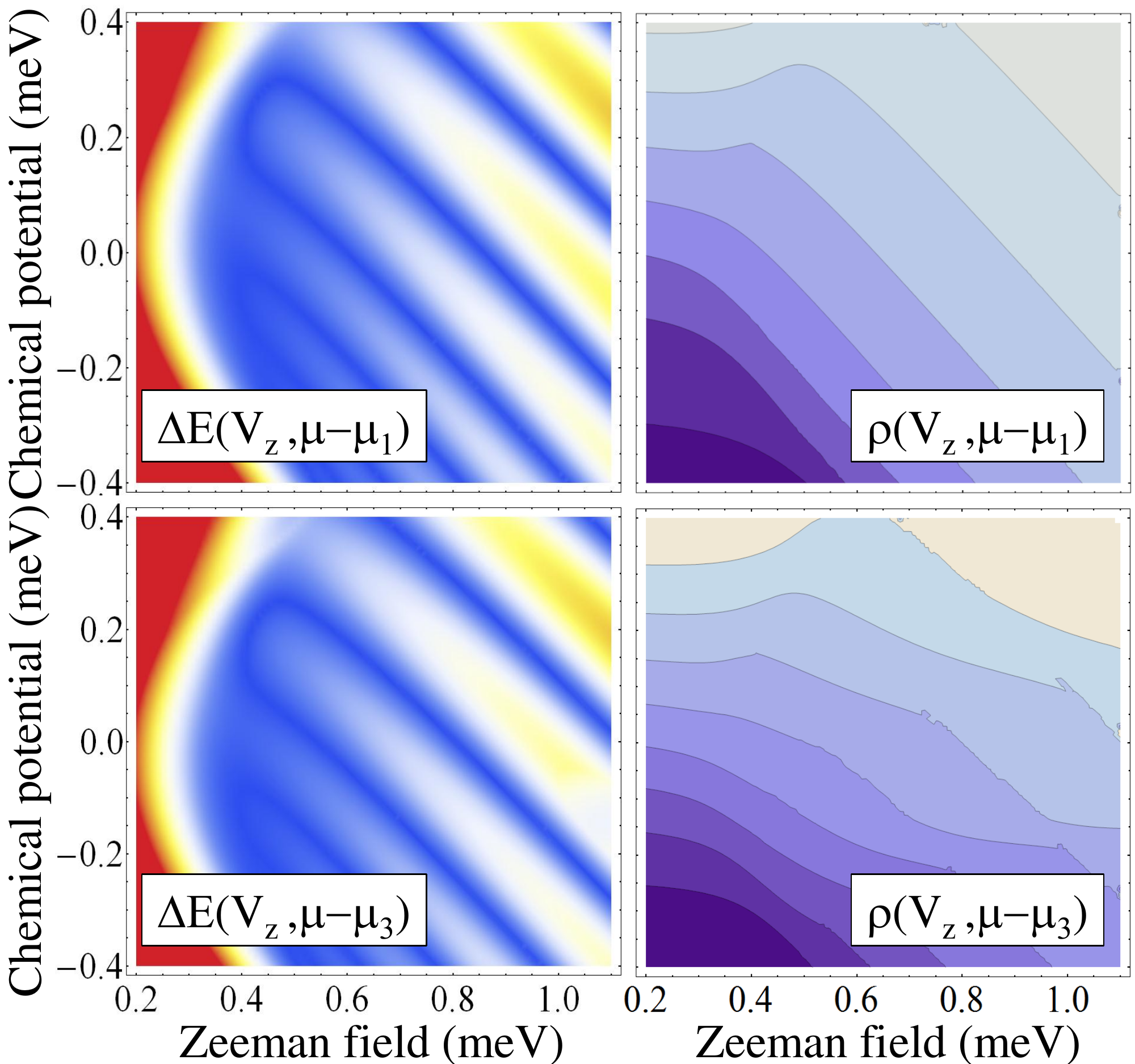}
\vspace{-7mm}
\end{center}
\caption{(Color online) Left panels: MF splitting as a function of chemical potential $\mu$ and Zeeman field $V_Z$ for single--band system (top) and for a wire with $n_{\rm top}=3$ (bottom). The dark blue regions correspond to the minima of the energy splitting.  The parallel bands represent the source of the oscillatory splitting and a hallmark for Majorana physics. Right panel: Density as a  function of chemical potential $\mu$ and Zeeman splitting $V_Z$.  Note that, in the single--band case,  the constant density contours are nearly parallel with the structures in the left panel, which may results in a strongly suppressed splitting or in a monotonic dependence of the splitting on $V_Z$.}
\vspace{-6mm}
\label{Fig4}
\end{figure}

In the more realistic (and necessarily more complex) multi--band case, the particle density is not tied to the effective  Fermi wave---vector of the top band, so one expects the oscillatory behavior to be partly restored. Nonetheless, the quasi--periodicity of the oscillatory splitting is not a generic feature in a wire with constant particle density. Specific examples are  shown in  the lower panels of Fig. ~\ref{Fig3} corresponding to a system with $n_{\rm top}=3$ (i.e., five occupied spin sub--bands). In systems with more occupied bands, the dependence of the splitting on $V_z$ approaches the quasi--periodic oscillatory behavior characteristic of the constant chemical potential case. To gain a deeper understanding of the manifestations of the Majorana oscillatory splitting in different conditions, we show in Fig. \ref{Fig4} (left panels) the dependence of $\Delta E$ on both the Zeeman field $V_Z$ and the chemical potential $\Delta\mu$, together with contour plots of the electron number as function of the same variables (right panels). 
 Examining Fig.~\ref{Fig4} we note that both the constant density contours and the characteristic parallel structures characterizing the MF splitting slope in  a similar direction. Therefore, the MF splitting as a function of the Zeeman field at constant density will show far slower oscillations, than the constant chemical potential situation. Moreover, a constant density path may lie within either a low--splitting or a high--splitting band for a significant range of Zeeman fields, thus explaining the types of behavior illustrated in Fig.~\ref{Fig3}.

Our work clearly establishes that the Majorana-induced zero bias conductance peak should manifest some oscillatory splitting at large applied magnetic fields (or as a function of the gate-tuned chemical potential at constant magnetic field), and a direct observation of this oscillatory splitting behavior could be the smoking gun evidence for the Majorana existence.

This work is supported by Microsoft Q, DARPA-QuEST and the Harvard Quantum Optics Center. 

\appendix

\section{Numerical methods}
 We consider a semiconductor nanowire with rectangular cross section and dimensions $L_x \gg L_y\sim L_z$. In the numerical calculations we have $L_y\approx 90$nm, $L_z\approx 60$nm, and two wire lengths,  $L_x =1\mu$m and $L_x =2\mu$m. For an $InSb$ nanowire with this cross section, the typical energy gap between the subbands is of the order $5$mev. We assume that only a few subband are occupied and we project the single particle quantum mechanics problem for the nanowire electrons onto a low--energy subspace spanned by the states with energies lower than a certain threshold (typically $\sim100$meV). As a result of the proximity to an s-wave superconductor (SC), a pair potential $\Delta$ is induced in the nanowire and the energy scale for the quantum states in the semiconductor (SM) are renormalized. To account for this proximity effect, we integrate out the SC degrees of freedom and incorporate them as a surface self--energy term of the form \cite{SLDS}
\begin{align}\label{eq:Sigma_clean}
\Sigma(\omega)=-\gamma\left[ \frac{\omega + \Delta_0 \sigma_y\tau_y}{\sqrt{ \Delta_0^2-\omega^2}}+\zeta \tau_z\right],
\end{align}
where $\gamma=0.3$meV is the effective SM-SC coupling, $\tau_x$ and $\tau_z$ are Pauli matrices in the Nambu space, $\Delta_0=1.5$meV is the pair potential of the bulk SC, and $\zeta$ is a proximity-induced shift of the chemical potential. In the present calculations we take $\zeta=0$. Within the static approximation $\sqrt{ \Delta_0^2-\omega^2} \rightarrow \Delta_0$, the self-energy becomes $\Sigma(\omega)\approx -\gamma\omega/\Delta_0  - \gamma \sigma_y\tau_y$
and the low-energy physics of the SM nanowire with proximity-induced SC can be described by an effective Bogoliubov-de Gennes Hamiltonian. 
This approximation is valid, strictly speaking, at energies much lower than $\Delta_0$, but represents a very good approximation even for $E\sim\Delta_0/2$. Explicitly, the matrix elements of the effective BdG Hamiltonian can be written as
\begin{eqnarray}
 H_{\rm BdG}({\bm n},{\bm n}^\prime) &=& Z \left[\epsilon_{\bm n} \delta_{{\bm n n}^\prime} + V_Z \sigma_x \delta_{{\bm n}{\bm n}^\prime} + \langle H_{\rm SOI}^x\rangle_{{\bm n n}^\prime} \right]\tau_z \nonumber \\
&+& Z \langle H_{\rm SOI}^y\rangle_{{\bm n n}^\prime} + \Delta \sigma_y \tau_y,  \label{Hbdg}
\end{eqnarray} 
where ${\bm n} =(n_x, n_y, n_z)$ are quantum numbers for the nanowire states in the absence of spin--orbit coupling, 
$\epsilon_{\bm n}$ are the corresponding energies, $V_Z$ is the Zeeman splitting, and $ \langle H_{\rm SOI}^x\rangle_{{\bm n n}^\prime}$ and $ \langle H_{\rm SOI}^y\rangle_{{\bm n n}^\prime}$  are matrix elements for the intra--band and inter--band  Rashba spin--orbit coupling, respectively. Note that energy scale for the SM nanowire is renormalized by a factor $Z= (1+\gamma/\Delta_0)^{-1}$ due to the SC proximity effect. This renormalization is determined by the term in the self-energy (\ref{eq:Sigma_clean}) that is proportional to $\omega$ (in the static approximation). The pairing term in Eq. (\ref{Hbdg}) is derived from the corresponding contribution to the self-energy (\ref{eq:Sigma_clean}) and is proportional to the induced pair potential $\Delta=\gamma\Delta_0/(\gamma+\Delta_0)=250\mu$eV. The effective Hamiltonian described by Eq. (\ref{Hbdg}) is diagonalized numerically.

\section{Analytic solution for Majorana wave-functions in nanowires}
The one-dimensional BdG Hamiltonian for a single band semiconductor nanowire
with spin-orbit coupling (assumed to be linear in the momentum $k_y$),
 which in general can be written as
\begin{equation}
H_{BdG}=(- \partial_x^2-\mu(x))\tau_z + V_z \sigma_z+\imath\alpha\partial_x\sigma_y\tau_z+\Delta\tau_x
\end{equation}
where $V_Z$ is the strength of the effective Zeeman field
and the unit vector $\alpha$ characterizes the  spin-orbit coupling. 

Non-degenerate Majorana spinor solutions are of the form $\Psi=(u,\imath\sigma_y u^*)$ and are completely determined
 by the 2-spinor $u$.  This fact
was used to obtain the Majorana solutions for vortices to reduce the
BdG equation from a $4\times 4$ system of equations to a $2\times 2$
system of equation. However, this reduction procedure required the BdG Hamiltonian
to be real which is not the case for general forms of spin-orbit coupling
and Zeeman splitting.
The BdG equation for the zero energy mode $H_{BdG}\Psi=0$
may be reduced to an equation for $u$ as
\begin{equation}
\left[(-\partial_x^2-\mu(x)) + V_Z\sigma_z+\alpha(\imath \sigma_x)\partial_x\right]u+\Delta(\imath\sigma_y)u^*=0.
\end{equation}
This equation  is not real but may be reduced to a system of real
 equations by
writing $u=u_R+\imath u_I$ and taking the real and imaginary parts of
the resulting equation giving a pair of equations of the form
\begin{align}
&\left[(-\partial_x^2-\mu(x)) + V_Z\sigma_z+\alpha(\imath \sigma_y)\partial_x+\Delta(\imath\sigma_y)\right]u_R=0\\
&\left[(-\partial_x^2-\mu(y)) + V_Z\sigma_z+\alpha(\imath \sigma_y)\partial_x-\Delta(\imath\sigma_y)\right]u_I=0.
\end{align}
This results in a reduced BdG equation for the $E=0$ reduced spinor
 $\Psi(x)$
\begin{equation}
\left(\begin{array}{cc}- \partial_x^2+V_z-\mu(x) & \lambda\Delta+\alpha  \partial_x\\ -\lambda \Delta-\alpha \partial_x  & - \partial_x^2-V_z-\mu(x) \end{array}\right)u(x)=0
\end{equation}
where $\lambda=\pm 1$.

An end  of the wire of the type considered above is defined by requiring some parameter of the Hamiltonian to vary across
 the end situated at $x=0$. We take this parameter to be constant for
 $x<0$ and $x>0$. In this case, our
previous approach can be applied in a way even simpler than the application to the vortex problem, since the solutions on both sides of the
interface at $x=0$ can be approximated as a sum 
\begin{equation}
u(x)=\sum_n a_n e^{-z_n x}\rho_n\label{eq:u}
\end{equation}
 where 
\begin{align}
&\left(\begin{array}{cc}- z_n^2 +V_z-\mu & \lambda\Delta-z_n\alpha  \\ -\lambda \Delta+z_n\alpha    & - z_n^2-V_z-\mu \end{array}\right)\rho_n=0.\label{eq:largeRmode_edge}
\end{align}
The solutions $z_n$ must satisfy the quartic equation 
\begin{align}
&Det\left(\begin{array}{cc}- z^2 +V_z-\mu & \lambda\Delta-z\alpha  \\ -\lambda\Delta+z\alpha    & - z^2-V_z-\mu \end{array}\right)
\nonumber\\
&=( z^2+\mu)^2-V_z^2+(z\alpha\mp \Delta)^2=0\label{eq:char_eq}.
\end{align}
For 
\begin{equation}
C_0=(\Delta^2+\mu^2-V_Z^2)<0,
\end{equation}
 there are 3 values of $z_n$ such that $Re(z_n)<0$ in a given
$\lambda$ channel.
The coefficients $a_n$ in the solution are determined
 by matching the boundary conditions on $\Psi(x)$ at $x=0$.

The equation Eq.~\ref{eq:char_eq}, while in principle has an exact solution, the solution 
is extremely complicated to write out explicitly. Therefore, we will discuss an 
approximate solution, which will be valid in the limit where either $\alpha$ or $\Delta$
are small. In this case Eq.~\ref{eq:char_eq} is written as 
\begin{align}
&( z^2+\mu)^2-V_z^2+(z\alpha\mp \Delta)^2\nonumber\\
&=( z^2+\mu)^2-V_z^2+(z^2\alpha^2\mp 2\alpha\Delta z+\Delta^2)=0\nonumber\\
&=z^4+(\alpha^2+2\mu)z^2-(V_z^2-\Delta^2-\mu^2\pm 2\alpha\Delta z)=0\nonumber\\
&z= \pm\nonumber\\
& \sqrt{-(\mu+\alpha^2/2)\pm\sqrt{(\mu+\alpha^2/2)^2+(V_z^2-\Delta^2-\mu^2\pm 2\alpha\Delta z)}}.
\end{align}
In the limit where spin-orbit coupling and $\alpha$ or $\Delta$ is small, the gap in the bulk BdG spectrum is small 
and one can have nearly propagating solutions, which are approximated as below.
Since the values of $z$ determine the $x$-dependence of the solutions of Eq.~\ref{eq:u}, 
$z$ can be written  as 
\begin{align}
&z=\pm i k_{F,eff}-\xi^{-1},\label{eq:kFxi}
\end{align}
where $\xi$ is the effective coherence length and $k_{F,eff}$ is the effective 
fermi-wave-vector associated with the zero-mode solution $\Psi(x)\propto e^{i k_{F,eff}x-x/\xi}$ for $x\gg\xi$.
In the limit where either of $\alpha$ or $\Delta$ are small, $\xi$ can be approximated by 
\begin{equation}
\xi^{-1}\approx\frac{\alpha\Delta}{2\sqrt{(\mu+\alpha^2/2)^2+(V_Z^2-\Delta^2-\mu^2)}}\label{eq:xi}
\end{equation}
and $k_{F,eff}$ can be approximated by 
\begin{equation}
k_{F,eff}\approx\sqrt{\sqrt{(\mu+\alpha^2/2)^2+(V_Z^2-\Delta^2-\mu^2)}+(\mu+\alpha^2/2)}.\label{eq:kFeff}
\end{equation}

To determine the full Majorana wave-function in Eq.~\ref{eq:u}, in addition to $z_n$ one needs to determine 
 $\rho_n=(u_{n,\ua},u_{n,\da})$. The spinors $\rho_n$ are calculated by solving the equation  
\begin{equation}\label{eq:z2}
\left(\begin{array}{cc}- z^2+V_z-\mu &\lambda\Delta- z\alpha  \\ -\lambda\Delta+z \alpha  & - z^2 -V_z-\mu \end{array}\right)\left(\begin{array}{c} u_{\uparrow} \\ u_{\downarrow}\end{array}\right)=0.
\end{equation}
The solutions are now given by 
\begin{align}
&u_{\uparrow}=-(\lambda\Delta- z\alpha)\\
&u_{\downarrow}=- z^2+V_z-\mu,
\end{align}
where $z$ is an acceptable solution for the quartic equation Eq.~\ref{eq:char_eq}.
 (Eq.~\ref{eq:char_eq}), for $V_z^2>(\Delta^2+\mu^2)$ there are 3 solutions on the right half of the complex $z$ plane and 1 solution on the left
 half for $\lambda=-1$. The situation is opposite for $\lambda=1$.
Since the equations are real, the 3 solutions are $z,z^*$ and $w$ where $w$ is real. The wave-function $\psi(x)$ is then written as 
\begin{align}
&u(x)=e^{-w x}\left(\begin{array}{c}-(\lambda\Delta- \rho\alpha)\\- \rho^2+V_z-\mu\end{array}\right) +a e^{-z x}\left(\begin{array}{c}-(\lambda\Delta- z\alpha)\\- z^2+V_z-\mu\end{array}\right)\nonumber\\
&+a^* e^{-z^* x}\left(\begin{array}{c}-(\lambda\Delta- z^*\alpha)\\- z^{* 2}+V_z-\mu\end{array}\right).
\end{align} 
The boundary condition $\psi(0)=0$ implies 
that 
\begin{align}
& a=\frac{i(z^*-w)[V_Z\alpha-z^*(\lambda \Delta-\alpha w)-(\alpha\mu+\Delta\lambda w) ]}{2 Im(z)(V_z\alpha+|z|^2\alpha-2 Re[z]\Delta\lambda-\alpha\mu)}.
\end{align}
Since the four components of the spinor are related by particle-hole symmetry, $\psi=(u,i\sigma_y u*)$, we can write $\psi$ as 
\begin{align}
&\psi(x)=e^{-w x}\left(\begin{array}{c}-(\lambda\Delta- w\alpha)\\- w^2+V_z-\mu\\- w^2+V_z-\mu\\(\lambda\Delta- w\alpha)\end{array}\right) +a e^{-z x}\left(\begin{array}{c}-(\lambda\Delta- z\alpha)\\- z^2+V_z-\mu\\- z^2+V_z-\mu\\(\lambda\Delta- z\alpha)\end{array}\right)\nonumber\\
&+a^* e^{-z^* x}\left(\begin{array}{c}-(\lambda\Delta- z^*\alpha)\\- z^{* 2}+V_z-\mu\\- z^{* 2}+V_z-\mu\\(\lambda\Delta- z^*\alpha)\end{array}\right).
\end{align} 

The left Majorana fermions $\psi_l(x)$ satisfies 
\begin{align}
&[(-\partial_x^2-\mu-i\partial_x \sigma_y)\tau_z+V_z \sigma_z+\Delta\tau_x]\psi_l(x)=0.
\end{align}
An inverse solution is obtained by noting that $\psi_r(x)=i \sigma_z\psi_l(-x)$ is a solution as well, which decays in the opposite direction.
The factor of $i$ ensures that $\psi_r(x)$ is invariant under $\sigma_y\tau_y K$.

\section{Splitting of MFs in spin-orbit coupled nanowires}
To compute Majorana fermion splitting we proceed using the variational principle by writing 
\begin{equation}
\Psi(x)=c_l\psi_l(x)+c_r\psi_r(x),
\end{equation}
where $\psi_{l,r}(x)$ are the Majorana fermion wave-functions, which are zero-modes of the Hamiltonian for $x>0$ and $x<L$ respectively that have been 
appropriately truncated at the other boundary. Here $\Psi(x)$ is expected to be a solution 
for a wire with chemical poetntial domain wall at both ends $x=0,L$. 
\begin{align}
&E=\frac{\int dx \Psi^\dagger(x)H\Psi(x)}{\int dx \Psi^\dagger(x)\Psi(x)}\nonumber\\
&=\frac{c_l^* c_r\int dx\psi_l^*(x)H\psi_r(x)+c_r^* c_l\int dx\psi_r^*(x)H\psi_l(x)}{|c_l|^2+|c_r|^2+c_l^* c_r\int dx\psi_l^*(x)\psi_r(x)+c_r^* c_l\int dx\psi_r^*(x)\psi_l(x)}\nonumber\\
&\approx c_l^* c_r\int dx\psi_l^*(x)H\psi_r(x)+c_r^* c_l\int dx\psi_r^*(x)H\psi_l(x)\nonumber\\
&\approx c_l^* c_r\int_0^{L/2} dx\psi_l^*(x)H\psi_r(x)+c_r^* c_l\int_0^{L/2} dx\psi_r^*(x)H\psi_l(x)\nonumber\\
&+ c_l^* c_r\int_{L/2}^L dx\psi_l^*(x)H\psi_r(x)+c_r^* c_l\int_{L/2}^L dx\psi_r^*(x)H\psi_l(x),
\end{align}
where we have assumed $|c_l|^2+|c_r|^2=1$ and also used the fact that $\psi_{l,r}(x)$ are particle-hole symmetric
 (so their expectation value for $H$ vanishes).
Observing that $H\psi_r(x)=0$ for $x>L/2$ and $H\psi_l(x)=0$ for $x<L/2$, 
\begin{align}
&E\approx c_l^* c_r\int_0^{L/2} dx\psi_l^*(x)H\psi_r(x)+c_r^* c_l\int_{L/2}^L dx\psi_r^*(x)H\psi_l(x).
\end{align}
Using Green theorem, 
\begin{align}
&\int_a^b dx \psi^*(x)\partial_x\phi(x)=\int_a^b dx \partial_x(\psi^*(x)\phi(x))-\partial_x\psi^*(x)\phi(x)\nonumber\\
&=[(\psi^*(x)\phi(x))]_a^b-\int_a^b dx \partial_x\psi^*(x)\phi(x)\\
&\int_a^b dx \psi^*(x)\partial_x^2\phi(x)=\int_a^b dx \partial_x(\psi^*(x)\partial_x\phi(x))-\partial_x\psi^*(x)\partial_x\phi(x)\nonumber\\
&=[(\psi^*(x)\partial_x\phi(x))]_a^b-\int_a^b dx \partial_x(\partial_x\psi^*(x)\phi(x))\nonumber\\
&+\int_a^b dx \partial_x^2\psi^*(x)\phi(x)\nonumber\\
&=[(\psi^*(x)\partial_x\phi(x))-\partial_x\psi^*(x)\phi(x)]_a^b+\int_a^b dx \partial_x^2\psi^*(x)\phi(x)\\
&\int_0^{L/2}dx \psi_l^*(x)[(-\partial_x^2-\mu+i\alpha \sigma_y \partial_x)\tau_z+V_z \sigma_z+\Delta\tau_x]\psi_r(x)\nonumber\\
&=\int_0^{L/2}dx [\{(-\partial_x^2-\mu+i \alpha \sigma_y \partial_x)\tau_z+V_z \sigma_z+\Delta\tau_x\}\psi_l^*(x)]\psi_r(x)\nonumber\\
&+[(\psi^*_l(x)\partial_x\tau_z\psi_r(x)+i\alpha \psi^*_l(x)\sigma_y\tau_z\psi_r(x) )]_0^{L/2}\nonumber\\
&=(\psi^*_l(L/2)\partial_x\tau_z\psi_r(L/2)+i\alpha \psi^*_l(L/2)\sigma_y\tau_z\psi_r(L/2) ).
\end{align}
Using the relation $\psi_r(x)=\sigma_z\psi_l(L-x)$, we note that the above matrix element becomes 
\begin{align}
&(\psi^*_l(L/2)\partial_x\tau_z \sigma_z\psi_l(L/2)+i\alpha \psi^*_l(L/2)\sigma_y\tau_z \sigma_z\psi_l(L/2) ).\label{eq:overlap}
\end{align}
In the case of a long wire in the case where one of the modes has a significantly longer decay length 
\begin{align}
&E\approx \hbar^2 k_F\frac{e^{-L/\xi}}{m\xi}\cos{(k_F L)},\label{eq:simple1}
\end{align}
where $k_F,\xi$ are given by Eq.~\ref{eq:xi} and Eq.~\ref{eq:kFeff}.
This equation appears as Eq.~\ref{eq:simple} in the main manuscript and is the central result of this 
section of the Supplementary material.
Based on Eq.~\ref{eq:kFeff}, $k_F$ depends on $V_Z$, which enters into a rapidly varying (in the limit of large $L$) 
cosine term in Eq.~\ref{eq:simple1}.
 Therefore, one can expect the splitting $E$ to oscillate as a function 
of $V_Z$ as discussed in the main text. 

\section{Role of Coulomb interactions - self-consistent calculations}
Let us now suppose that the electrons in the wire are subject to a Coulomb interaction that is screened by 
the superconducting substrate by perfect metallic screening.  Within the Hartree (the Fock term is ignored here) 
approximation, one expects a renormalization of the external potential 
\begin{equation}
V_{ext}(x)=\int dx' K(x-x')\rho(x'),
\end{equation} 
where $K(x-x')$ is the metallic screened Coulomb kernel and $\rho(x')$ is the density of electrons on the nanowire.
Within BCS theory, the total charge density $\rho(x')$ is given by 
\begin{equation}
\rho(x)=\sum_{n:E_n<0,\sigma}|u_n(x,\sigma)|^2,
\end{equation}
where $n$ is summed over states with $E_n<0$. As a consistency check note that in the absence of superconductivity 
$\int dx \sum_{\sigma}|u_n(x,\sigma)|^2=1$ for states with $(\epsilon_n-\mu)<0$ and vanishes for other states.
Therefore $\int dx\rho(x)$ is the total number of electrons for $\Delta=0$.

For simplicity, we ignore the spatial variation of $\rho(x)$ and approximate 
\begin{equation}
\rho(x)\approx \rho 
\end{equation} 
and also ignore the variation in $V_{ext}(x)$ so that 
\begin{equation}
V_{ext}(x)\approx V_{ext}=\kappa^{-1}\rho, 
\end{equation}
where $\kappa$ is the mean compressibility.

The mean-external potential $V_{ext}$ renormalizes $\mu$ in the BdG equation according to 
\begin{equation}
\mu\rightarrow \tilde{\mu}=\mu-V_{ext}.
\end{equation}
Assuming that the compressibility is small $\kappa\rightarrow 0$, it is clear that a small change of 
$\rho$ changes $V_{ext}$ and therefore $\tilde{mu}$ by a large amount. This leads to the constraint 
that when parameters other than $V_{gate}$ are changed 
\begin{equation}
\delta \int dx\rho(x)\approx 0.
\end{equation}
It is possible to change $\rho(x)$ by changing the gate voltage - but as in mesoscopic wires, one 
needs to change $V_{gate}$ by many volts before a change in density corresponding to a few meV change 
of chemical potential is accomplished.

In the simple single-band case and ignoring superconductivity $\Delta$, the linear density of electrons 
is given by $n=k_F/2$ and the chemical potential $\mu$ is given by 
\begin{equation}
\mu=n^2/4-\sqrt{V_Z^2-n^2\alpha^2/4}.
\end{equation} 
Substituting this expression for $\mu$ into Eq.~\ref{eq:kFeff} we find that in the limit $\Delta\rightarrow 0$, 
$k_{F,eff}\approx n/2$, which is independent of $V_Z$. 
Since based on Eq.~\ref{eq:simple}, the oscillations as a function of  $V_Z$ are a result of the $V_Z$ dependence 
of $k_{F,eff}$, the oscillations in the $V_Z$ dependence can be expected to be suppressed in the constant density 
limit. This is confirmed in the single-band case by results in Fig.~\ref{Fig1}.

The multi-band results are more complicated and in principle can depend on the inter and intra band capacitances of 
various channels. However, assuming that all the capacitances are the same, the multi-band effect is expected to 
screen the role of the coulomb interaction, so that changing $V_z$ leads to some change of $k_F$. The result (shown in Fig.~\ref{Fig1}) 
leads to more $V_Z$-dependence of the splitting than in the single-band case. 

The $V_Z$ dependence of the splitting in the multi-band constant density case can be understood by examining Fig.~\ref{Fig2}.


\begin{thebibliography}
\expandafter\ifx\csname natexlab\endcsname\relax

\fi
\expandafter\ifx\csname bibnamefont\endcsname\relax

\fi
\expandafter\ifx\csname bibfnamefont\endcsname\relax

\fi
\expandafter\ifx\csname citenamefont\endcsname\relax

\fi
\expandafter\ifx\csname url\endcsname\relax

\fi
\expandafter\ifx\csname urlprefix\endcsname\relax

\fi
\providecommand{\bibinfo}[2]{#2} \providecommand{\eprint}[2][]{\url{#2}}

 \bibitem{Mourik} V. Mourik, K. Zuo,  S. M. Frolov, S. R. Plissard, E. P. A. M. Bakkers and L. P. Kouwenhoven, Science DOI: 10.1126/science.1222360.


\bibitem{Sau} Jay D. Sau, R. M. Lutchyn, S. Tewari, S. Das Sarma,
 Phys. Rev. Lett. \textbf{104}, 040502 (2010).

 \bibitem{Long-PRB} J. D. Sau, S. Tewari, R. Lutchyn, T. Stanescu and S. Das Sarma, Phys. Rev. B \textbf{82}, 214509 (2010).

 \bibitem{Roman} R. M. Lutchyn, J. D. Sau, S. Das Sarma, Phys. Rev. Lett. \textbf{105}, 077001 (2010).

 \bibitem{Oreg} Y. Oreg, G. Refael, F. V. Oppen, Phys. Rev. Lett. \textbf{105}, 177002 (2010).



\bibitem{Reich}E. S. Reich, Nature (2012).

\bibitem{Wilczek}F. Wilczek, Nature \textbf{486}, 195 (2012).

\bibitem{Physicstoday}R. Mark Wilson, Physics Today, \textbf{65}, 14 (2012).


\bibitem{Brouwer_science}  P. Brouwer, Science, \textbf{336},  989 (2012).




\bibitem{Deng} M. T. Deng, C. L. Yu, G. Y. Huang, M. Larsson, P. Caroff, H. Q. Xu, arXiv:1204.4130.

\bibitem{Weizman}A. Das, Y. Ronen, Y. Most, Y. Oreg, M. Heiblum, H. Shtrikman, arXiv:1205.7073.
\bibitem{Marcus} C. M. Marcus, private communication (2012).

\bibitem{Sengupta-2001} K. Sengupta, I. Zutic, H.-J. Kwon, V. M. Yakovenko, S. Das Sarma, Phys. Rev. B \textbf{63}, 144531 (2001).

\bibitem{R1}
 K. T. Law, Patrick A. Lee, and T. K. Ng, Phys. Rev. Lett.
\textbf{103}, 237001 (2009); K. Flensberg, Phys. Rev. B \textbf{82}, 180516 (2010);
 M. Wimmer, A.R. Akhmerov, J.P. Dahlhaus, C.W.J. Beenakker, New J. Phys. \textbf{13}, 053016 (2011).



\bibitem{Prada}
    E. Prada, P. San-Jose, R. Aguado, arXiv:1203.4488 (2012).

\bibitem{Stanescu2012}
    T. D. Stanescu, S. Tewari, J. D. Sau, S. Das Sarma, arXiv:1206.0013 (2012).

\bibitem{Lin2012}
    C-H Lin, J. D. Sau, S. Das Sarma,  arXiv:1204.3085 (2012).



\bibitem{Liu} J. Liu, A. C. Potter, K.T. Law, P. A. Lee, arXiv:1206.1276 (2012).

\bibitem{Altland}     D. Bagrets, A. Altland, arXiv:1206.0434 (2012).

\bibitem{Beenakker_Weak} D. I. Pikulin, J. P. Dahlhaus, M. Wimmer, H. Schomerus, C. W. J. Beenakker, arXiv:1206.6687 (2012).

 \bibitem{Kells} G. Kells, D. Meidan, P. W. Brouwer, arXiv:1207.3067 (2012).


\bibitem{Silvano}
    E. J. H. Lee, X. Jiang, R. Aguado, G. Katsaros, C. M. Lieber, S. De Franceschi, arXiv:1207.1259 (2012).

\bibitem{kitaev} A. Y. Kitaev, Physics-Uspekhi \textbf{44}, 131 (2001).
\bibitem{Kwon} H.-J. Kwon, K. Sengupta, V. M. Yakovenko, The European Physical Journal B \textbf{37}, 349-361 (2004).


\bibitem{alicea}J. Alicea, Y. Oreg, G. Refael, F. von Oppen, M. P. A. Fisher, Nature Physics \textbf{7}, 412-417 (2011);
J. D. Sau, D. J. Clarke, S. Tewari, Phys. Rev. B 84, 094505 (2011); 
B. van Heck, A.R. Akhmerov, F. Hassler, M.Burrello, C.W.J. Beenakker, New J. Phys. \textbf{14}  035019 (2012).

\bibitem{sau_int}F. Hassler, A. R. Akhmerov, C.-Y. Hou, and C. W. J.
Beenakker, New J. Phys. \textbf{12}, 125002 (2010);
J. D. Sau, S. Tewari, S. Das Sarma, Phys. Rev. A 82, 052322 (2010);
 L. Fu, Phys. Rev. Lett. 104, 056402 (2010);
J. Sau, B. Swingle, S. Tewari, arXiv:1210.5514 (2012).


\bibitem{2D_int} L. Fu, C. Kane, Phys. Rev. Lett. 
 \textbf{\bibinfo{volume}{102}},
  \bibinfo{eid}{216403} (\bibinfo{year}{2009}); A. Akhmerov, J. Nilsson, C. Beenakker,
  \bibinfo{journal}{Phys.\ Rev.\ Lett.} \textbf{\bibinfo{volume}{102}},
  \bibinfo{eid}{216404} (\bibinfo{year}{2009}); 
J. D. Sau, S. Tewari, S. Das Sarma, Phys. Rev. B \textbf{84}, 085109 (2011); 
E. Grosfeld, B. Seradjeh, S. Vishveshwara, Phys. Rev. B \textbf{83},
 104513 (2011); E. Grosfeld, A. Stern, Proc. Natl. Acad. Sci. U S A, \textbf{108}, 11810-4 (2011). 


\bibitem{meng}M. Cheng, R. M. Lutchyn, V. Galitski, S. Das Sarma,
Phys. Rev. Lett. \textbf{103}, 107001 (2009);M. Cheng, R. M. Lutchyn, V. Galitski, S. Das Sarma
, Phys. Rev. B \textbf{82}, 094504 (2010). 

\bibitem{SLDS} T. D. Stanescu,  R. M. Lutchyn, S. Das Sarma, Phys. Rev. B {\bf 84}, 144522 (2011).


\end{thebibliography}
\end{document}